# Asymmetric solitons and domain walls supported by inhomogeneous defocusing nonlinearity

Yaroslav V. Kartashov,[1,2] Valery E. Lobanov,[1] Boris A. Malomed,[3] and Lluis Torner[1]

[1]*ICFO-Institut de Ciencies Fotoniques, and Universitat Politecnica de Catalunya, 08860 Castelldefels (Barcelona), Spain*
[2]*Institute of Spectroscopy, Russian Academy of Sciences, Troitsk, Moscow Region, 142190, Russia*
[3]*Department of Physical Electronics, School of Electrical Engineering, Faculty of Engineering, Tel Aviv University, Tel Aviv 69978, Israel*

We show that an inhomogeneous defocusing nonlinearity that grows toward the periphery in the positive and negative transverse directions at different rates can support strongly asymmetric fundamental and multipole bright solitons, which are stable in wide parameter regions. In the limiting case when nonlinearity is uniform in one direction, solitons transform into stable domain walls (fronts), with constant or oscillating intensity in the homogeneous region, attached to a tail rapidly decaying in the direction of growing nonlinearity.

OCIS Codes: 190.4360, 190.6135

Interest in the evolution of light beams in materials with spatially inhomogeneous parameters, such as refractive index or nonlinearity, is motivated by the possibility to control diffraction broadening for beam shaping and steering. For example, the strength and sign of the effective diffraction can be *managed* in periodic refractive-index landscapes, with the aim to create nonlinear modes and propagation regimes that are not possible in uniform media [1-3]. Modulation of the local strength of the nonlinearity can be also used to control the beam dynamics via effective *pseudo-potentials* whose impact on light propagation crucially depends on its intensity. Various types of solitons were predicted in nonlinear pseudo-potentials [4], including one-dimensional solitons in nonlinear [5-9] and combined linear-nonlinear [10-12] lattices, exact modes supported by specially designed localized focusing nonlinearities [13], two-dimensional solitons supported by localized or periodic nonlinearities [14-17]. Formation of bright solitons in pseudo-potentials requires the presence of domains with focusing nonlinearity, assuming that the nonlinearity modulation depth is limited. However, it was recently shown that purely defocusing nonlinearities also support bright solitons with a finite total power, provided that the nonlinearity strength grows toward the periphery of the medium faster than $r^D$, where $D$ is the spatial dimension [18-21]. An unusual property of such solitons is that their symmetry and asymptotic form are determined by the nonlinearity profile, and do not depend on the propagation constant. Thus far, only symmetric solitons were predicted in settings of this type, while neither strongly asymmetric modes nor their limit form representing domain walls (DWs), which separate filled and empty domains, have been discovered. Such DWs have been studied only in materials with linear-refractive-index landscapes [22-25].

In this Letter we show that asymmetric defocusing nonlinearities characterized by different rates of the nonlinearity growth in the positive and negative transverse directions, can give rise to asymmetric bright solitons that may be stable. They turn into stable DWs if the nonlinearity becomes uniform in one direction.

We consider the propagation of light along the $\xi$ axis in a medium with a spatially inhomogeneous defocusing nonlinearity governed by the nonlinear Schrödinger equation for the dimensionless light-field amplitude, $q$:

$$i\frac{\partial q}{\partial \xi} = -\frac{1}{2}\frac{\partial^2 q}{\partial \eta^2} + \sigma(\eta) q |q|^2, \qquad (1)$$

where $\eta$ is the transverse coordinate and $\sigma(\eta) > 0$ is the local strength of the defocusing nonlinearity. Here we consider a spatially inhomogeneous nonlinearity growing toward the periphery as $\sigma(\eta<0) = \exp(\alpha_l \eta^2)$ and $\sigma(\eta \geq 0) = \exp(\alpha_r \eta^2)$, where $\alpha_{l,r}$ define the nonlinearity growth rates at $\eta<0$ and $\eta>0$, respectively. We set $\alpha_r = 1$ by means of rescaling and vary $\alpha_l$, which results in asymmetric nonlinearity distributions. Such nonlinearity profiles may be realizable by inhomogeneous doping of suitable photorefractive materials, or by the inhomogeneous application of Feshbach resonances in the case of matter waves [4].

Soliton solutions of Eq. (1) with propagation constant $b$ are looked for as $q(\eta,\xi) = w(\eta)\exp(ib\xi)$. Their stability was investigated by adding small perturbations $u(\eta), v(\eta)$, $q = [w + u\exp(\delta\xi) + iv\exp(\delta\xi)]\exp(ib\xi)$, and linearizing Eq. (1), which leads to the eigenvalue problem $\delta u = -(1/2)d^2v/d\eta^2 + bv + \sigma w^2 v$, $\delta v = (1/2)d^2u/d\eta^2 - bu - 3\sigma w^2 u$ for the instability growth rate $\delta$.

Despite the defocusing character of the nonlinearity, we have found a variety of asymmetric bright soliton solutions of Eq. (1), which are classified by the number of zeros (nodes) $k$ in their shapes. Figure 1 depicts representative examples. The existence of such states is a consequence of the *nonlinearizability* of Eq. (1) for the decaying tails of the solitons, due to the growing nonlinearity strength [18,19]. When $\alpha_l$ decreases solitons develop a wide left lobe, as solutions adapt to the slower rate of the nonlinearity growth at $\eta<0$.

Several noteworthy observations are suggested by Fig. 1. First, for a given value of $\alpha_l$, soliton solutions with different values of $k$ feature identical asymptotic forms at $\eta \to \pm\infty$ as follows from the comparison of fundamental and dipole solitons in Fig. 1(a), although their shapes differ considerably around $\eta=0$. Second, for different nonlinearity growth rates $\alpha_1$ at $\eta<0$, the soliton tails tend to coincide at $\eta>0$, as seen in the plot. Further, the soliton width rapidly grows when $\alpha_1$ decreases, due to the appearance of a wide left lobe, and the width diverges at $\alpha_1 \to 0$, with the nodes appearing in the profile of multipole solitons simultaneously shifting to $\eta<0$.

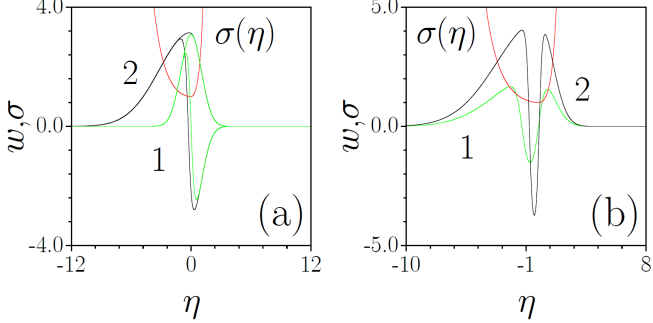

Fig. 1. (Color online) (a) Stable fundamental and dipole solitons at $b=-10$ for $\alpha_1=1$ (curves 1) and $\alpha_1=0.1$ (curves 2). (b) Stable tripole solitons at $\alpha_1=0.1$ for $b=-5$ (curve 1) and $b=-20$ (curve 2). Nonlinearity landscapes $\sigma(\eta)$ are shown for $\alpha_1=0.1$.

Solitons exist with negative values of the propagation constant, which is explained by the fact that the stationary version of Eq. (1) yields $b<0$ at the inflexion point, $\partial^2 w/\partial \eta^2=0$. The soliton amplitude increases with increasing $-b$ [Fig. 1(b)], consistent with the Thomas-Fermi approximation applied to this case [18]. While the asymptotic form of the solutions at $\eta \to \pm \infty$ is solely determined by $\alpha_1$ and does not depend on $b$, the soliton core changes considerably with $b$. In particular, increasing $|b|$ pushes all nodes to the right [Fig. 1(b)].

The soliton width, defined as $W=2U^{-1}\int_{-\infty}^{\infty}|\eta||q|^2\,d\eta$, where $U=\int_{-\infty}^{\infty}|q|^2\,d\eta$ is the total energy flow, rapidly decreases by increasing $|b|$, saturating at $b<-20$. Higher-order solutions, with a larger number of nodes, always carry a smaller energy flow [Fig. 2(a)]. On the other hand, the energy flow grows when $\alpha_1$ decreases and diverges at $\alpha_1 \to 0$, simultaneously being a monotonically increasing function of $|b|$ [Fig. 2(b)]. Dependencies similar to those shown in Figs. 2(a) and 2(b) were also obtained for modes with a larger number of nodes.

Linear stability analysis indicates that there is a broad range of parameters $(b,\alpha_1)$, where solutions are stable for all values of the number of nodes $k$ up to $k=10$, at least. In particular, fundamental and dipole solitons are stable in the entire existence domain. Tripoles feature two instability domains, which are indicated by gray areas in Fig. 2(c). Remarkably, stability is possible at $\alpha_1 \to 0$ too, when the soliton strongly expands to $\eta<0$. Inside the gray areas in Fig. 2(c), the instability of tripoles is oscillatory, resulting in irregular shape oscillations, as seen in Fig. 3(a). In contrast to systems with homogeneous nonlinearity, where instability-induced emission of radiation may be considerable, in the setting analyzed here almost all light stays around the nonlinearity minimum, even if the unstable beam exhibits considerable oscillations. The structure of the stability and instability domains becomes more complex as $k$ increases, but stability domains are always found.

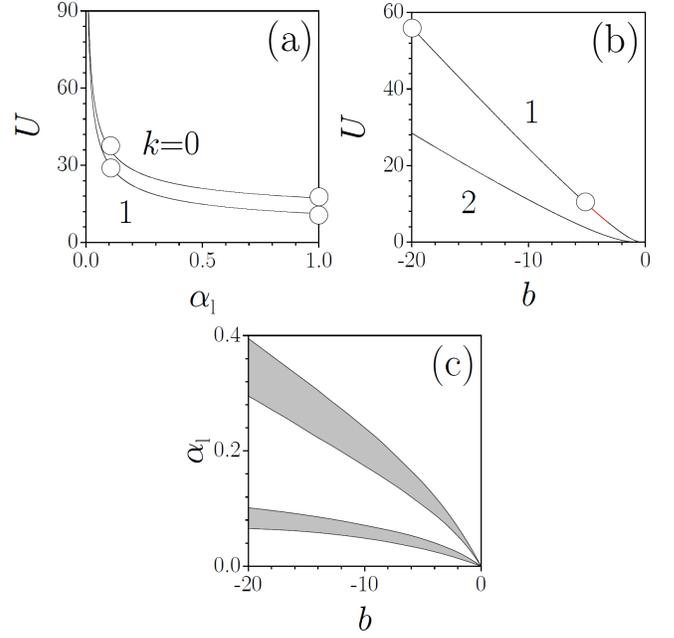

Fig. 2. (Color online) (a) Energy flow versus $\alpha_1$ for fundamental ($k=0$) and dipole ($k=1$) solitons at $b=-10$. Circles correspond to solitons in Fig. 1(a). (b) Energy flow versus $b$ for tripole ($k=2$) solitons at $\alpha_1=0.1$ (curve 1) and $\alpha_1=0.4$ (curve 2). A short red segment of curve 1 correspond to unstable tripoles. Circles correspond to solitons in Fig. 1(b). (c) Domains of stability (white) and instability (shaded) for tripole solitons.

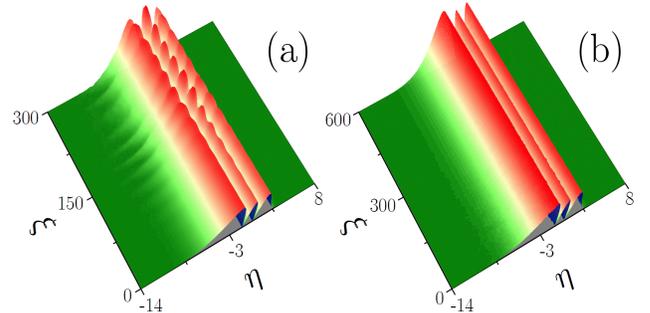

Fig. 3. (Color online) Evolution of unstable (a) and stable (b) tripoles with $b=-4$ and $b=-6$, respectively, and $\alpha_1=0.1$.

At $\alpha_1 \to 0$, the soliton solutions transform into DWs separating the decaying tail at $\eta>0$ and a cnoidal wave at $\eta \to -\infty$, whose amplitude for a given propagation constant $b$ takes the values $w_0 \leq w_{\max}=|b|^{1/2}$. In fact, $w_0$ is the second free parameter of the DW solutions, with the maximal amplitude $w_{\max}$ corresponding to an asymptotically flat state, as shown in Fig. 4. Note that solutions of this type can be found in an exact form for $b=-3/2$ and $w_0^2=(3-7^{1/2})/2$:

$$w(\eta)=\begin{cases} 2^{-1/2}\eta\exp(-\eta^2/2), & \text{at } \eta\geq 0, \\ w_0\text{sn}(\beta\eta,\kappa), & \text{at } \eta<0, \end{cases} \quad (2)$$

where $\beta^2=(3+7^{1/2})/2$ and $\kappa^2=(3-7^{1/2})/(3+7^{1/2})$ determines the elliptic modulus of the sn function.

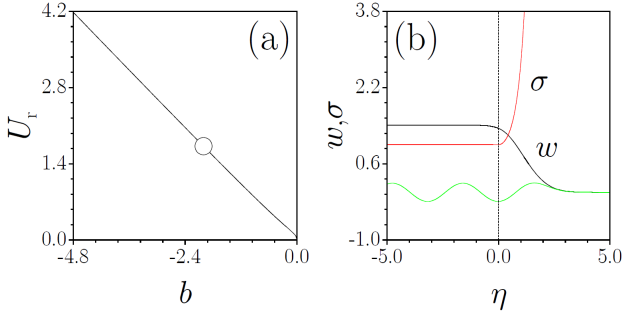

Fig. 4. (a) (Color online) Renormalized energy flow of the domain walls versus the propagation constant, for $\alpha_1=0$. (b) The profile of the domain wall with $b=-2$ (the black line) corresponding to the circle in (a). The solution with the cnoidal-wave form at $\eta<0$ is shown by the green line, also for $b=-2$.

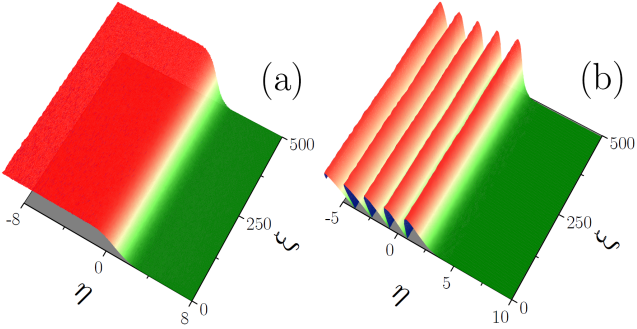

Fig. 5. (Color online) Stable propagation of a domain wall (a) and solution with the cnoidal-wave form at $\eta<0$ (b) in the presence of initial noise. Both solutions correspond to $b=-2$.

To the best of our knowledge, this is first known example of domain walls supported exclusively by an inhomogeneous defocusing nonlinearity, without the help of an additional modulation of the linear refractive index. The domain walls with constant amplitude $w_0=w_{\max}$ at $\eta\to-\infty$ may be characterized by the renormalized energy flow, $U_r=\int_{-\infty}^{\infty}|w^2-\theta(\eta)w_0^2|d\eta$, where $\theta(\eta)=1$ for $\eta<0$ and $0$ for $\eta\geq 0$, which is a monotonously decreasing function of $b$, as shown in Fig. 4(a). A detailed stability analysis performed on the basis of the linear eigenvalue problem shows that DWs with constant amplitude are stable for all values of $b$. This conclusion is supported by direct simulations of the propagation of perturbed DWs, which maintain their shapes over indefinitely long distances [Fig. 5(a)]. Their counterparts with cnoidal-wave asymptotic form,s including the exact solution (2), also appear to be stable in the simulations, as illustrated in Fig. 5(b), but the full analysis of their stability is a challenging numerical problem. Note that an exact solution of the latter type can be found for another nonlinearity modulation profile, namely, $\sigma(\eta>0)=\cosh^2\eta$ and $\sigma(\eta<0)\equiv 1$, in the form

$$w(\eta)=\begin{cases} 3^{1/2}\sinh(\eta)\operatorname{sech}^2(\eta), \text{ at } \eta\geq 0, \\ w_0\operatorname{sn}(\beta\eta,\kappa), \text{ at } \eta<0, \end{cases} \quad (3)$$

with squared amplitude $w_0^2=(5-13^{1/2})/2$, the elliptic modulus given by $\kappa^2=(5-13^{1/2})/(5+13^{1/2})$, and $\beta^2=(5+13^{1/2})/2$. This solution corresponds to $b=-5/2$.

Summarizing, we have shown that defocusing nonlinear media with a nonlinearity strength increasing toward the periphery at different rates in two contiguous regions support strongly asymmetric stable fundamental and multipole solitons. Solitons are stable in large parameter regions, and they transform into stable domain walls. The results reported here may be generalized to the case of two transverse dimensions.